# Conceptual Design of LiFi Audio Transmission Using Pre-Programmed Modules

Auwal Tijjani Amshi
auwalamshi@gmail.com

*Abstract-* We all know that Wi-Fi is presently the most commonly used technology for data transmission and connecting devices to the Internet, at the same time due to much reasonable concern, (such as Wi-Fi can be vulnerable when it comes to hacking, health concern, and low latency, etc.) the concept of Li-Fi is becoming very popular as a new way of data transmission that use light waves to transmit data rather than radio waves. Light-emitting diodes LED are used when transmitting the data in the visible light spectrum. Li-fi uses visible light communication and it has a promising future. Unlike Wi-fi, Li-Fi has low latency, high efficiency, accessible spectrum, and high data can be achieved. It is highly secured so the data cannot be hacked. In this paper, we design a concept of Li-fi audio signal transmission by reusing and repurposing pre-programmed modules to simplify and discuss visible light communication (VLC) in other to give a new researcher the idea on how the concept of LiFi and VLC. In addition to designing the concept we experiment to test the concept and we illustrated the result within this paper.

## 1 Introduction

Transferring data from one place to another is one of the most important day-to-day activities. The current wireless networks that connect us to the internet are very slow when multiple devices are connected.[1,2] As the number of devices that access the internet increases, the fixed bandwidth available makes it more and more difficult to enjoy high data rates and connect to a secure network.[2]

Everyone is now interested in using his mobile phone, laptop to communicate with others through Wireless-Fidelity (Wi-Fi) systems, and this technology, Wi-Fi, is frequently used among people in all public locations such as homes, shops, resorts, and airports, and the usage of wireless systems is also increasing rapidly every year; but the capacity is reducing Given the limitations of Radio Frequency (RF) resources, we're going to have a significant challenges. [3]

In order to solve this challenge in the longer term, this new Light-Fidelity technology (Li-Fi). Li-Fi is a short-range wireless system that's also useful for data transmission via Led bulbs. It uses visible light, a part of the electromagnetic spectrum that is still not greatly utilized, instead of the RF part[4,5].

The contribution of this paper will include designing a concept that can illustrate the feasibility of applying LiFi technology in underwater communication. Another basic contribution to this paper is to bring the explanation of LiFi Technolgy and VLC to the level where anybody with a basic understanding of engineering can understand the stated technology.

## 2 Visible light communication (VLC)

Visible light communication (VLC) VLC is an optical communication technology that uses visible light rays, located between [400-800] THz, as an optical carrier for data transmission by illumination. It uses fast electrical pulses that cannot be perceived by the human eye to transmit information [6]. One of VLC"s features is providing wide bandwidth as illustrated in Figure 1. We can see that using the optical portion of the spectrum guarantees about 10,000 times greater bandwidth compares to the usage of the RF frequencies[7].

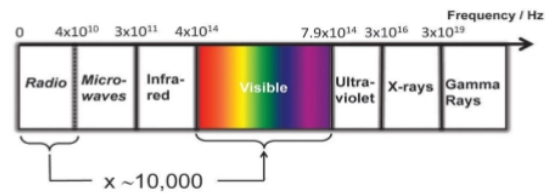

Figure 1. Location Of Visible Light And Rf Frequencies At Electromagnetic Spectrum[5].

### 2.1 Introduction Light–Fidelity (Li-Fi)

LiFi technology consists of LED as the media transmission and photo-detector as a receiver of transmitted data. A lamp driver is needed to make the LED work properly.[8] While amplification and processing are responsible to manage the signal that comes from the photo-detector.

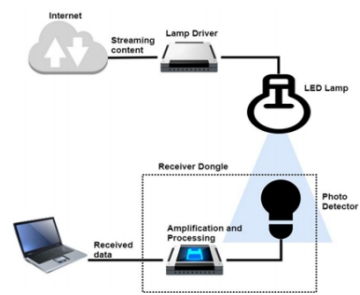

Figure 2. Basic Concept Diagram LiFi

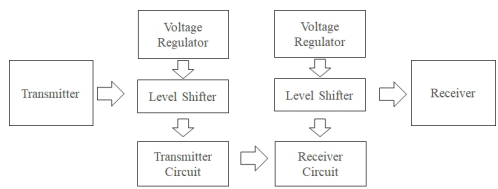

Figure 3. The working principle of the Li-Fi

Figure 2 shows the working principle of the Li-Fi system, for data transmission; it can be done by a single LED or multi LED. On the receiver side, there is a photo-detector, which converts this light into electric signals and it will give the electric signals to the device connected to it. Voltage regulator and level shifter circuits are used on both sides to convert or maintain a voltage level between transmitter and receiver.

## 2.2 Comparison Between LI-FI and WI-FI

The efficiency and safety of the internet are the dominating issues now. LiFi was found in 2011 by Scientist Harold Haas from the UK. The design is to overcome the disadvantage of WiFi. The speed of WiFi is up to 1500mbps and it's not sufficient to accommodate a huge user.[8,9,10,11]

**Table 1**. Basic Difference between LiFi and WiFi

| COMPARISON | LI-FI | WI-FI |
|---|---|---|
| Full form | Stand for light fidelity | Stands for wireless fidelity |
| Invented/coined | Coined by prof. Harald Haas in 2011 | By NCR corporation in 1991 |
| Operation | It transmits data using light with the help of led bulbs | It transmits data using radio waves using a wifi router |
| Technology | Present IrDA compliant devices | Wlan 802.11/b/g/n/ac/d standard compliant devices |
| Data transfer speed | About 1 Gbps | Ranges from 150mbps to maximum of 2gbps |
| Privacy | Light is blocked by the walls hence provide more secure data transfer | Walls cannot block radio waves so we need to employ more techniques to achieve secure data transfer |
| Frequency of operation | 10, 000 times frequency spectrum of the radio | 2.4ghz, 4.9ghz, and 5ghz |
| Coverage distance | About 10 meters | About 32 meters(vary based on transmit power and antenna type) |
| Data density | Work with the high dense environment | Work in a less dense environment due to interference related issues |
| Bare minimum components used | Led bulb, led driver, and photo-detector | Routers, modems, and access points |
| Applications | Used in airlines, undersea exploration, etc | Used for internet browsing with the help of wifi hotspot |

## 3 VLC System Design

Like all kinds of communication, this system consists of the transmission section and receiver section. In the transmission section, we have only three components, the LED, Power, and the auxiliary AUX. meanwhile, in the receiving section, we have a solar panel, a speaker, a power source, and an audio amplifier.

### 3.1 Transmission Section

In the transmission section, there is a LED, a battery, and an auxiliary cable (AUX) for audio transmission. The AUX is also known as audio jack receive signals form device and transfer analog audio signals to the LED. Which drives the LED using on-off-keying (OOK) Modulation. If the LED is on, it transmits a digital 1 otherwise it transmits a digital 0. The LEDs are switched on and off quickly to transmit data that can't be detected by a human eye.

### 3.2 LI-FI Transmitter Concept One

As a First concept, we build the transmitter in a way that signals can be received from a given device directly using an AUX Cable then the signal controls the transistor which opens and close the power supply to the LED. The transmitter block diagram is shown in Figure 4.

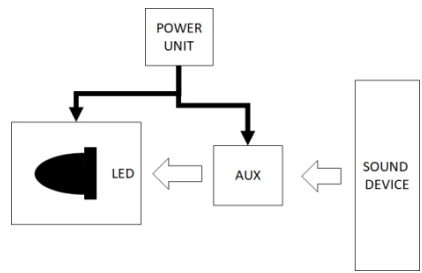

**Figure 4.** Transmitter Block Diagram.

### 3.2.1 Light-Emitting Diodes (LED)

LEDs are just tiny light bulbs that fit easily into an electrical circuit. But unlike ordinary bulbs, they are illuminated solely by the movement of electrons in a semiconductor material. The light intensity can be manipulated to send data by tiny changes in amplitude.[12]

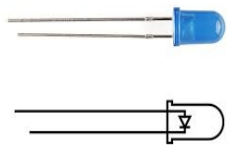

Figure 5. Light-Emitting Diodes (LED)

As previously discussed in the paper Li-Fi relies on the modulation of light at frequencies that are imperceptible to the human eye in most lighting applications. As semiconductors, LEDs can turn on and off up to a million times per second, enabling the diodes to send data quickly [13].

### 3.2.2 Auxiliary cable (AUX)

In this study, we use an auxiliary cable (AUX) which is a standard communications cable. AUX is an asynchronous serial cable with an interface that allows the auxiliary input of audio signals for MP3 players, Headphones, Portable music players, Amplifiers, Speakers[14].

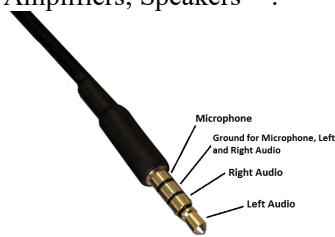

Figure 6. AUX Cable.

The AUX port is typically used for audio equipment that receives peripheral sound sources, such as digital music players or audio speakers. The peripheral sound device is connected to an AUX port or other medium such as a vehicle's audio jack.

### 3.3 LI-FI Transmitter Concept Two

As the second concept, we build the transmitter in a way that it will receive signals from a given device using a Bluetooth module in other to make the transmission somewhat completely wireless. Section 3.3.1 explains more about the Bluetooth module.

### 3.3.1 Bluetooth Amplifier Module.

XY-bt-mini Bluetooth is mainly used for Bluetooth voice players. It supports Bluetooth wireless connections. Smart home playback.

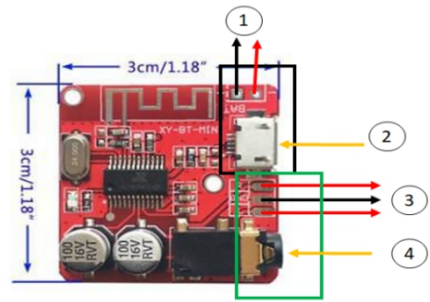

Figure 7. Bluetooth Amplifier

From diagrams 1 and 2 are power input. 3 and 4 are for communication (transmission). 2 and 4 are USB Port and AUX port. 1 and 3 are alternative to 2 and 4.

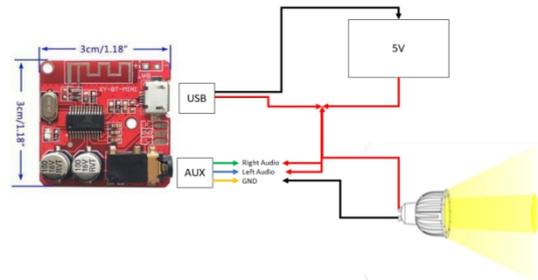

Figure 8. Using the module AUX Port to connect AUX Communication Cable and USB Cable to connect the Power Source.

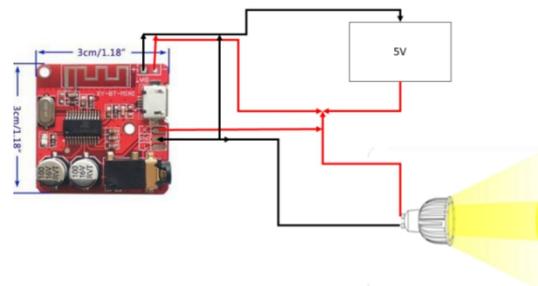

Figure 9. Directly connecting the module to Power and LED.

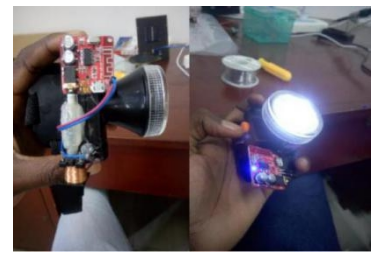

Figure 10. Transmitter Module

## 3.4 LI-FI Transmitter Concept Three

As the third concept, we added a Microphone Sound Detection Sensor module to replace the audio source (device). Section 3.4.1 discuss further on the module.

### 3.4.1 Microphone Sound Detection Sensor Module (KY-038).

KY-038 is a microphone sound sensor, which detects sound. It gives a measurement of how loud a sound is. [15]The sensor modules have a built-in potentiometer to adjust the sensitivity of the digital output pin.

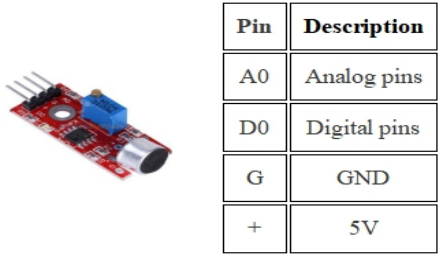

| Pin | Description |
|-----|-------------|
| A0  | Analog pins |
| D0  | Digital pins |
| G   | GND |
| +   | 5V |

Figure 11. KY-038 microphone sound sensor

The module has four pins, pin DO pin AO, pin +, and pin G.

## 3.5 Receiver Section

On the receiver section, there is a solar panel, which converts this light into electric signals and it will pass the electric signals to the PAM8403 audio amplifier.

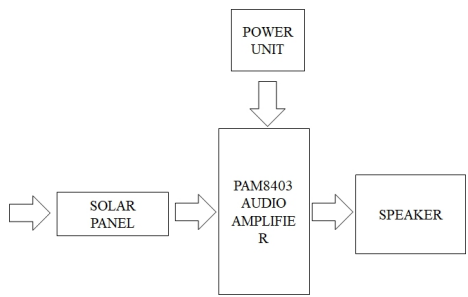

Figure 12. Receiver Block Diagram

### 3.5.1 Li-Fi Receiver

The photo-detector in this case the solar panel absorbed the ones and zeros from the LED source and These signals (ones and zeros) are demodulated and amplified by the audio amplifier. The light intensity is absorbed by the solar panel then the audio signal comes out using a speaker.

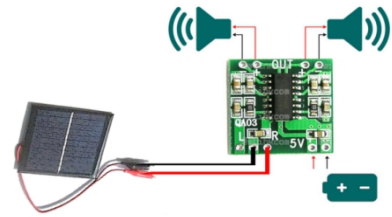

Figure 13. Receiver circuit diagram.

### 3.5.2 Audio Amplifier (PAM8403)

The PAM8403 is a 3W, class-D audio amplifier. PAM8403 can amplify sound that is given from the AUX OR microphone. This circuit can apply to LCD Monitors / TV Projectors, Notebook Computers, Portable Speakers, Portable DVD Players, Game Machines, Cellular Phones/Speaker Phones. it is inexpensive and low power consumption and only needs a little component to work. [16]

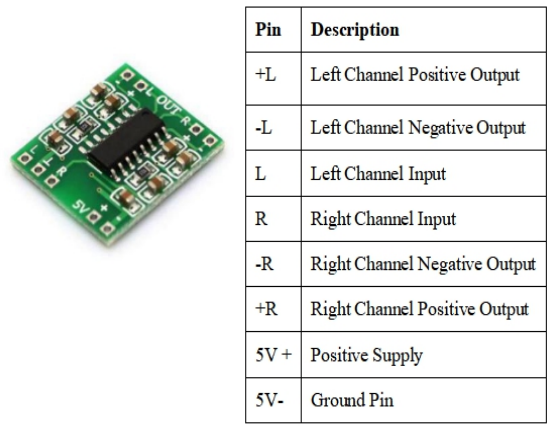

| Pin | Description |
|-----|-------------|
| +L  | Left Channel Positive Output |
| -L  | Left Channel Negative Output |
| L   | Left Channel Input |
| R   | Right Channel Input |
| -R  | Right Channel Negative Output |
| +R  | Right Channel Positive Output |
| 5V + | Positive Supply |
| 5V- | Ground Pin |

Figure 14. Typical PAM8403 Applications Circuit.

With the same numbers of external components, the efficiency of the PAM8403 is much better than that of Class-AB cousins. Moreover, PAM8403 is suited for portable applications because it extends battery life.

### 3.5.3 Speaker

In this study, we use a speaker that converts the electrical or analog signal to the audible form to reach the receptor. The audio receptor receiver is the output/message sand by the transmitter.

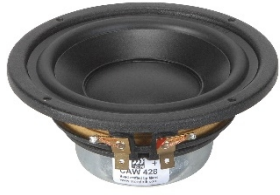

Figure 15. Speaker

### 3.5.4 Solar Panel

We used the solar panel as a broadband receiver, the encoded audio data will be fed from LED driver to LED, which will further pass the data in the digitized form to the solar cells. Solar cells are self-sufficient in the sense they will capture energy as well as transmit data.

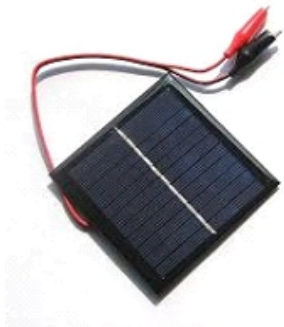

Figure 16. Solar Panel.

Moreover, the system also consists of a solar concentrator, which is placed above the solar panel. This will increase the intensity of the solar pane.

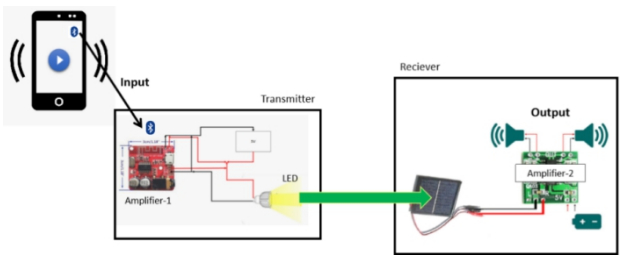

Figure 17. Propose LiFi Concept Two.

## 4 Experimentation

After the design of the concept, we did a simple experiment, where we used a module called Arduino and python serial programming to print and plot the data received from our transmitter by our receiver module. Fig 18 illustrate the first 500 received audio signal point.

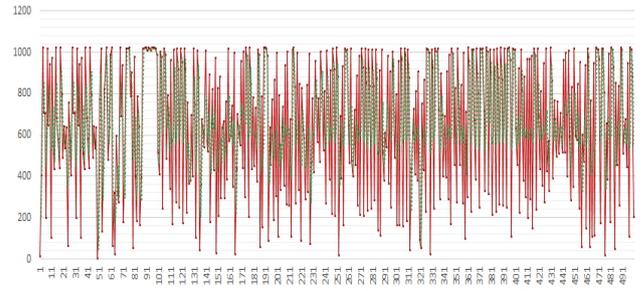

Figure 18. 1st 500 Audio signal Trend line.

From the figure above we can observe that two signal trend lines are overlapping each other red and green. The red one represents the original signal point received by our receiver while the green one represents the average movement of the signal Trend line. The main reason for plotting the above graph is to give an idea of how the signal received by our proposed LiFi receiver looks like without any work (such as noise reduction) done on the signal data.

## 5 Conclusion

In this paper, we proposed a real-time high-quality audio signal transmission and receiving using reprogrammable modules. Transmitting distance is about 10-15m can be achieved. Because Li-Fi provides secure, low cost, easy data transfer, and reliable communication, it can be used in industrial, medical, and military applications. Arduino and python serial are used for successful transmission testing.

## 6 Future Scope

The future scope of this project will include an improved concept which will include the image, text transmission, internet data transmission, etc. Also, we will introduce noise filtering and further discuss the real-time application of LiFi at home, institutions, industries, and many more areas.